# Maximum correntropy criterion based sparse adaptive filtering algorithms for robust channel estimation under non-Gaussian environments

Wentao Ma[a], Hua Qu[a], Guan Gui[b*], Li Xu[b], JihongZhao[a], BadongChen[a*]

[a]School of Electronic and Information Engineering, Xi'an Jiaotong University, Xi'an, 710049, China

[b]Department of Electronics and Information Systems, Akita Prefectural University, 015-0055, Japan

**Abstract:**

Sparse adaptive channel estimation problem is one of the most important topics in broadband wireless communications systems due to its simplicity and robustness. So far many sparsity-aware channel estimation algorithms have been developed based on the well-known minimum mean square error (MMSE) criterion, such as the zero-attracting least mean square (ZALMS),which are robust under Gaussian assumption. In non-Gaussian environments, however, these methods are often no longer robust especially when systems are disturbed by random impulsive noises. To address this problem, we propose in this work a robust sparse adaptive filtering algorithm using correntropy induced metric (CIM) penalized maximum correntropy criterion (MCC) rather than conventional MMSE criterion for robust channel estimation. Specifically, MCC is utilized to mitigate the impulsive noise while CIM is adopted to exploit the channel sparsity efficiently. Both theoretical analysis and computer simulations are provided to corroborate the proposed methods.

**Keywords:**

Sparse channel estimation; sparse adaptive filtering; maximum correntropy criterion (MCC); correntropy induced metric (CIM); impulsive noise.

# 1. Introduction

Signal transmission over fading channel is often deteriorated severely. Hence, accurate channel estimation is one of the most important technical issues for realizing dependable wireless communications. In narrowband or wideband communications systems, wireless channels are often described in time domain by dense channel model, i.e., most of channel taps are nonzero coefficient [1].Linear channel estimation methods perform very well [1] to achieve the dense-model based lower bound. In broadband wireless communication systems, wireless channel in time domain often exhibits sparse structure which is supported by very few dominant coefficients while most of channels taps are zeros [2].The performance bound could be further achieved lower if they can exploit the sparsity in channels. Motivated by this fact, several effective sparse channel estimation (SCE)methods have been proposed to take advantage of channel sparsity and improve the estimation performance particularly in environments with high signal-to-noise ratio (SNR) [3-6]. However, these SCE methods may be instable under low SNR regimes. The potential problem motivates the fast development of sparse adaptive filtering algorithms which can work stable in low SNR regimes. In [7], proportionate-type algorithms were proposed to estimate channel parameters, which updates each channel coefficient in proportion to its estimated magnitude to exploit the sparse prior information. In [8], an improved proportionate normalized least mean square(IPNLMS) algorithm was proposed. In [9], a family of subband IPNLMS algorithms was proposed as well. In all above proportionate-type algorithms, the finite impulse response position to position and is roughly proportional at each tap position to the absolute value of the current tap weight estimate.

Besides, sparse adaptive algorithms were proposed by incorporating a sparsity-aware penalty term (e.g. $\ell_0$-norm or $\ell_1$-norm) into a traditional adaptive filtering algorithm such as the least mean square (LMS) or recursive least squares (RLS), motivated by the least absolute shrinkage and selection operator (LASSO) [10] and recent progresses in compressive sensing [11]. Well-known examples include zero-attracting LMS (ZALMS) [12], reweighted zero-attracting LMS (RZALMS) [12] and sparse RLS [13], which have been successfully applied in sparse system identification and sparse channel estimation. By adding different sparsity-aware penalty terms to the popular minimum mean square error (MMSE) criterion, these methods are very effective for sparse signal processing, such as sparse system identification [12] and sparse channel estimation [14].Many variants of sparse adaptive filtering algorithms have also been developed. For example, Wu et. al. proposed a novel sparse adaptive algorithm by introducing a variable $\ell_p$-norm-like constraint into the cost function of the LMS algorithm [15]. Aliyu et.al proposed a sparse variable step-size LMS algorithm by employing a $\ell_p$-norm constraint[16]. To take full advantage of channel sparsity, Gui et.al proposed several improved adaptive sparse channel estimation methods using $\ell_p$-norm normalized LMS($\ell_p$-NLMS) as well as $\ell_0$-norm normalized LMS ($\ell_0$-NLMS) [17]. In some severe deterioration scenarios, sparse least mean square (LMS)/fourth (LMF) algorithms (combined LMS and LMF algorithms) were proposed to improve performance as well as to exploit channel sparsity [18].

The above channel estimation methods were developed under Gaussian noise model. Accurately, as the development of channel measurement techniques, non-Gaussian noise models (e.g. alpha-stable noise model) were proposed to describe real communication environments which are more accurate than conventional model [19-22]. Under the non-Gaussian (especially heavy-tailed) noise environments, however, the existing methods are very sensitive to impulsive noises. To ensure more dependable signal transmission, robust channel estimation methods are necessary to mitigate the non-Gaussian impulsive noises. Recently, several robust adaptive algorithms were developed based on the maximum correntropy criterion (MCC) for signal processing [23-26]. The MCC is based on a new similarity measure called correntropy, which has some desirable properties [27]:1) it is always bounded foranydistribution;2) it contains all even-order moments, and the weights of the higher-order moments are determined by the kernel size(or kernel width); 3) it is a local similarity measure and is robust to outliers(more detailed theoretical analysis results can be found in [23]). The MCC based adaptive filtering algorithms may perform well with lower steady-state excess MSE (EMSE) [24]. So far correntropy has been applied in many areas. Singh et.al proposed a novel cost function for linear adaptive filtering by using correntropy[25]. Zhao et.al derived a robust kernel adaptive filter under MCC criterion [26].Yuan and Hu proposed a robust feature extraction framework based on correntropy[28]. He et.al. introduced a sparse correntropy framework for deriving robust sparse representations of face images for recognition [29–30].

To the best of our knowledge, however, it has not yet been proposed to design a robust and sparse adaptive filter to estimate a sparse channel in the presence of impulsive noises. In this work, we will develop a MCC based robust and sparse adaptive filtering algorithm to mitigate the non-Gaussian impulsive noises. In our approach, correntropy is used as a cost function to replace the traditional MSE, so that the resulting sparse filter is rather robust against the impulsive noises. In addition, we also introduce a new sparse penalty term to exploit the channel sparsity. It is well known that a sparse adaptive filtering algorithm using a stronger sparse penalty can exploit the channel prior information more efficiently [31].The SCE can be viewed as a sparse representation problem. However, finding the sparsest solution, which leads to an $\ell_0$-norm minimization problem, is an NP-hard combinatorial optimization problem. To deal with this intractable problem, some approximates of $\ell_0$-norm functions are usually used, such as the $\ell_1$-norm, reweighted $\ell_1$-norm. It is shown that the correntropy induced metric (CIM) [32] as a nonlinear metric in the input space can provide a nice approximation for the $\ell_0$-norm function. The goal of this paper is to develop a robust and sparse adaptive filtering algorithm by combining the MCC with the CIM. Specifically, the MCC is utilized to mitigate the impulsive noises while the CIM function imposes a zero attraction of the filter coefficients according to the relative value of each coefficient among all the entries which in turn leads to an improved performance when the system is sparse. Hence, our studies can be efficiently applied in sparse channel estimation under impulsive noise environments. Finally, numerical and simulation results are provided to corroborate the study.

The rest of the paper is organized as follows. In section 2, the correntropy and CIM are briefly reviewed. In section 3, the sparse MCC algorithm, namely CIMMCC, is derived. In section 4, the mean and mean-square convergence is analyzed. In section 5, simulation results are given to illustrate the desirable performance of the proposed algorithm. Finally, the work is summarized in section 6.

## 2. Correntropy and CIM

The correntropy is a nonlinear measure of the similarity between two random variables $X = [x_1, \cdots, x_N]^T$ and $Y = [y_1, \cdots, y_N]^T$ in kernel spaces [24]:

$$V(X,Y) = E[\kappa(X,Y)] = \int \kappa(x,y) dF_{XY}(x,y) \tag{1}$$

Where $E[\cdot]$ denotes the expectation operator, $\kappa(\cdot,\cdot)$ is a shift-invariant Mercer kernel, and $F_{XY}(x,y)$ denotes the joint distribution function. In practice, the data distribution is usually unknown, and only a finite number of samples $\{x_i, y_i\}$ are available. In this case, the correntropy can be estimated as:

$$\hat{V}(X,Y) = \frac{1}{N}\sum_{i=1}^{N}\kappa(x_i, y_i) \tag{2}$$

A typical kernel in correntropy is the Gaussian kernel:

$$\kappa(x,y) = \frac{1}{\sigma\sqrt{2\pi}}\exp(-\frac{e^2}{2\sigma^2}) \tag{3}$$

where $e = x-y$, and $\sigma$ denotes the kernel width. The correntropy is always bounded for any distribution and is robust to impulsive noises (or outliers) [23-24, 27]. The MCC criterion aims at maximizing the correntropy between a variable and its estimator, and this principle has been successfully applied in robust adaptive filtering [25-26].

The sparsity penalty term is a key factor in a sparse filter. The zero-attracting (ZA)and reweighted zero-attracting(RZA)penalty terms are usually adopted to design different sparse adaptive filtering algorithms [12-18, 33]. In this paper, we will introduce a CIM based sparse penalty to develop an efficient sparse MCC algorithm. For a better understanding, below we briefly revisit the CIM [23]. Consider $X$ and $Y$, in a sample space, the CIM is defined by

$$CIM(X,Y) = \left(\kappa(0) - \hat{V}(X,Y)\right)^{1/2}, \tag{4}$$

where $\kappa(0) = 1/(\sigma\sqrt{2\pi})$. The $\ell_0$-norm of the vector $X = [x_1,\cdots,x_N]^T$ can be approximated by [31]

$$\|X\|_0 \sim CIM^2(X,0) = \frac{\kappa(0)}{N}\sum_{i=1}^{N}\left(1-\exp(-x_i^2/2\sigma^2)\right). \tag{5}$$

It has been shown that if $|x_i| > \delta$, $\forall x_i \neq 0$, then as $\sigma \to 0$, we can get arbitrarily close to the $\ell_0$-norm [32], where $\delta$ is a small positive number. Due to its relation with correntropy, this nonlinear metric is called the correntropy induce metric. CIM is a nonlinear metric in the input space.

According to the above discussion, the CIM provides a well approximation for the $\ell_0$-norm. Hence,

it favors sparsity and can be used as a sparsity penalty term to exploit the channel sparsity in the sparse channel estimation scenarios.

## 3 CIMMCC algorithm

First of all, adaptive channel estimation problem is formulated as follows. An input signal vector $X(n) = [x_n, x_{n-1}, \cdots, x_{n-M+1}]^T$ is sent over a finite impulse response (FIR) filter channel with parameter vector $W_o = [w_{o,1}, w_{o,2}, \cdots, w_{o,M}]^T$, where $M$ is the size of the channel memory. It is assumed that the channel parameters are real-valued with sparse structure, i.e., most of channel coefficients are zeros. At the receive side, received signal $d(n)$ is obtained as

$$d(n) = W_o^T X(n) + v(n) \tag{6}$$

where $v(n)$ denotes the additive noise. In many existing systems, the noise $v(n)$ is described as non-Gaussian due to the impulsive nature of man-made electromagnetic interference as well as nature noise[36-40]. In order to mitigate the impulsive noises, we construct a cost function by combining MCC and CIM as follows

$$\begin{aligned} J_{CIMMCC}(W(n)) &= -J_{MCC}(W(n)) + \lambda J_{CIM}(W(n)) \\ &= -\frac{1}{\sigma_1 \sqrt{2\pi}} \exp(-\frac{e^2(n)}{2\sigma_1^2}) + \lambda \frac{1}{M\sigma_2 \sqrt{2\pi}} \sum_{i=1}^{M}\left(1 - \exp(-w_i(n)^2/2\sigma_2^2)\right), \end{aligned} \tag{7}$$

where $e(n)$ denotes the $n$-th update instantaneous error, i.e., $e(n) = d(n) - W^T(n)X(n)$, $W(n) = [w_1(n), w_2(n), \cdots, w_M(n)]^T$ stands for the $n$-th channel estimate; $\sigma_1$ and $\sigma_2$ represent the kernel widths of MCC and CIM, respectively. In (7), the MCC term plays robust role to impulsive noise while the CIM term (with smaller width) plays the role to exploit channel sparsity balanced by a weight factor $\lambda \geq 0$ which is a regularization parameter to balance between MCC estimation error and CIM sparsity.

Based on the cost function (7), a gradient based adaptive filtering algorithm can be derived as follows:

$$w_i(n+1) = w_i(n) - \eta \frac{\partial J_{CIMMCC}(n)}{\partial w_i(n)}$$

$$= w_i(n) - \eta \left[ \frac{-1}{\sigma_1^3 \sqrt{2\pi}} \exp\left(-\frac{e^2(n)}{2\sigma_1^2}\right) e(n) x_i(n) + \lambda \frac{1}{M\sigma_2^3 \sqrt{2\pi}} w_i(n) \exp\left(-\frac{w_i(n)^2}{2\sigma_2^2}\right) \right] \quad (8)$$

$$= w_i(n) + \mu \exp\left(-\frac{e^2(n)}{2\sigma_1^2}\right) e(n) x_i(n) - \rho \frac{1}{M\sigma_2^3 \sqrt{2\pi}} w_i(n) \exp\left(-\frac{w_i(n)^2}{2\sigma_2^2}\right).$$

The matrix-vector form of (8) can be also written as

$$W(n+1) = W(n) + \mu \exp(-e^2(n)/2\sigma_1^2) e(n) X(n) - \rho \frac{1}{M\sigma_2^3 \sqrt{2\pi}} W(n) \exp\left(-W(n)^2/2\sigma_2^2\right), \quad (9)$$

where $\mu = \eta/(\sigma_1^3 \sqrt{2\pi})$, and $\rho = \eta\lambda$. The algorithm in (9) is termed as CIMMCC. About the algorithm, it is worth noting that suitable selection of the kernel width can make the CIM approach the $\ell_0$-norm [23, 32]. Hence, the proposed algorithm can be applied in real systems. For a better understanding, the proposed algorithm is summarized as Table 1.

Table 1. The proposed CIMMCC algorithm

| Input parameters $\mu$, $\rho$, $\sigma_1$, $\sigma_2$, $M$ |
|---|
| Initial $W(0), X(0), d(0), e(0)$ <br> For $n=1,2,\ldots$ Do <br>     Input new $X(n)$ and $d(n)$ <br>     $e(n) = d(n) - W^T(n)X(n)$ <br>     $W(n+1) = W(n) + \mu \exp(-e^2(n)/2\sigma_1^2) e(n) X(n)$ <br>            $- \rho \frac{1}{M\sigma_2^3 \sqrt{2\pi}} W(n) \exp(-W(n)^2/2\sigma_2^2)$ <br> End |

**Remark 1:** The $\ell_1$-norm and reweighted $\ell_1$-norm are well-known popular sparsity penalty terms for constructing adaptive sparse filtering algorithms. Similarly, onecan also derive sparse MCC algorithms, e.g., $\ell_1$-norm into MCC (i.e., ZAMCC) and reweighted $\ell_1$-norm MCC (i.e., RZAMCC). The detailed derivation could be found in Appendix 1.

**Remark 2:** Compared with the traditional MCC algorithm, the computational complexity of the proposed CIMMCC algorithm is still very low. The operations per iteration of the CIMMCC contain 3M additions and 2M multiplications, and the exponential calculation is M+1.

## 4 Performance analysis

It is well known thatthe mean and mean square convergence are two key properties for an adaptive filter which is the fundamental of its feasibility. The convergence analysis of the proposed algorithm is in general a challenging problem. By means of an approximation approach, we study the mean and the mean square convergence of the proposed algorithm by using the generalization approximation idea [16, 33-35].Hence, the proposed algorithm can be rewritten as

$$W(n+1) = W(n) + \mu f(e(n))e(n)X(n) + \rho g(n) \tag{10}$$

where $f(e(n)) = \exp(-e^2(n)/2\sigma_1^2)$, and $g(n)$ is $-1/\left(M\sigma_2^3\sqrt{2\pi}\right)W(n)\exp(-W^2(n)/2\sigma_2^2)$ for the CIMMCC. To simplify the analysis, the following statistical assumptions are given.

*Assumption 1:* The input signal $\{X(n)\}$ is independent and identically distributed (i.i.d.) with zero-mean Gaussian distribution.

*Assumption 2:* The noise signal $\{v(n)\}$ is i.i.d. with zero-mean and variance $\sigma_v^2$, and is independent of $\{X(n)\}$.

*Assumption3:* The error nonlinearity $f(e(n)) = \exp(-e^2(n)/2\sigma_1^2)$ is independent of the input signal $\{X(n)\}$.

*Assumption4:* The $\{W(n)\}$ and $g(n)$ are independent of the $\{X(n)\}$.

*Assumption5:* The expectation $E[f(\mathrm{e}(\infty))]$ is limited.

**Remark 3:** Assumptions 1 and 2 are commonly used in [16, 33]. Assumption 3 is valid when the weight vector $W(n)$ lies in the neighborhood of the optimal solution $W_o$.

*4.1 Mean performance*

The filter misalignment vector is defined as

$$\tilde{W}(n) = W^* - W(n) \tag{11}$$

The mean and autocorrelation matrix of $\tilde{W}(n)$ are denoted by

$$\delta(n) = E[\tilde{W}(n)] \tag{12}$$

$$S(n) = E[\Delta(n)\Delta^T(n)] \tag{13}$$

where $\Delta(n)$ is

$$\Delta(n) = \tilde{W}(n) - \delta(n) = \tilde{W}(n) - E[\tilde{W}(n)] \tag{14}$$

Combining (6), (10) and (11), we obtain

$$\begin{aligned}
\tilde{W}(n+1) &= \tilde{W}(n) - \mu f(e(n))e(n)X(n) - \rho g(n) \\
&= \tilde{W}(n) - \mu f(e(n))(d(n) - W^T(n)X(n))X(n) - \rho g(n) \\
&= \tilde{W}(n) - \mu f(e(n))(W_o^T X(n) + v(n) - W^T(n)X(n))X(n) - \rho g(n) \\
&= \tilde{W}(n) - \mu f(e(n))X(n)X^T(n)\tilde{W}(n) - \mu f(e(n))v(n)X(n) - \rho g(n) \\
&= [I - \mu f(e(n))X(n)X^T(n)]\tilde{W}(n) - \mu f(e(n))v(n)X(n) - \rho g(n) \\
&= A(n)\tilde{W}(n) - \mu f(e(n))v(n)X(n) - \rho g(n)
\end{aligned} \tag{15}$$

where $A(n) = [I - \mu f(e(n))X(n)X^T(n)]$. Taking the expectation of (15) and using the independence assumptions 1, 2, and 3, we get:

$$\delta(n+1) = E[\tilde{W}(n+1)] = [1 - \mu E[f(e(n))]\sigma_x^2]\delta(n) - \rho E[g(n)] \tag{16}$$

where $\sigma_x^2$ denotes the covariance matrix of $X(n)$. From the (16), one can easily derive

$$\delta(\infty) = E[\tilde{W}(\infty)] = -\frac{\rho}{\mu E[f(e(\infty))]\sigma_x^2} E[g(\infty)] \tag{17}$$

Combining (11) and (17), we have

$$E[W(\infty)] = W^* - \frac{\rho}{\mu E[f(e(\infty))]\sigma_x^2} E[g(\infty)] \tag{18}$$

From the definition of $g(n)$ for the CIMMCC, we show that $E[g(\infty)]$ is still limited (see Appendix 2). Then under Assumption 5, $E[W(\infty)]$ is a bounded vector. As a result, $E[W(n)]$ will converge to a vector $E[W(\infty)]$ as shown in (18).

*4.2 Mean square performance*

Subtracting (16) from (15) and applying (14) yields

$$\begin{aligned}
\Delta(n+1) &= I - \mu f(e(n))X(n)X^T(n)]\tilde{W}(n) - \mu f(e(n))v(n)X(n) - \rho g(n) \\
&\quad -[1-\mu E[f(e(n))]\sigma_x^2]\delta(n) - \rho E[g(n)] \\
&= I - \mu f(e(n))X(n)X^T(n)]\tilde{W}(n) - [I - \mu f(e(n))X(n)X^T(n)]E[\tilde{W}(n)] \\
&\quad +[I - \mu f(e(n))X(n)X^T(n)]E[\tilde{W}(n)] - \mu f(e(n))v(n)X(n) \\
&\quad -\rho g(n) - [1-\mu E[f(e(n))]\sigma_x^2]\delta(n) - \rho E[g(n)] \\
&= I - \mu f(e(n))X(n)X^T(n)]\Delta(n) + \mu[E[f(e(n))]\sigma_x^2 - f(e(n))X(n)X^T(n)]\delta(n) \\
&\quad -\mu f(e(n))v(n)X(n) - \rho(g(n) - E[g(n)]) \\
&= A(n)\Delta(n) + \mu B(n)\delta(n) - \mu f(e(n))v(n)X(n) - \rho C(n)
\end{aligned} \quad (19)$$

where

$$B(n) = E[f(e(n))]\sigma_x^2 - f(e(n))X(n)X^T(n)$$

$$C(n) = g(n) - E[g(n)]$$

Under Assumptions 1-3, it is straightforward to verify that $B(n)$, and $C(n)$ are zero-mean, and $\tilde{W}(n)$, $X(n)$ and $v(n)$ are mutually independent. So, substituting (19) into (13) and after some tedious calculations, we can derive

$$\begin{aligned}
S(n+1) &= E[\Delta(n+1)\Delta^T(n+1)] \\
&= E\begin{bmatrix}(A(n)\Delta(n) + \mu B(n)\delta(n) - \mu f(e(n))v(n)X(n) - \gamma C(n)) \times \\ (A(n)\Delta(n) + \mu B(n)\delta(n) - \mu f(e(n))v(n)X(n) - \gamma C(n))^T\end{bmatrix} \\
&= \begin{bmatrix}E[A(n)\Delta(n)\Delta^T(n)A^T(n)] - \rho E[C(n)\Delta^T(n)A^T(n)] \\ +\mu^2 E[B(n)\delta(n)\delta^T(n)B^T(n)] + \mu^2 E[f^2(e(n))]\sigma_x^2\sigma_v^2 \\ -\rho E[A(n)\Delta(n)C^T(n)] + \rho^2 E[C(n)C^T(n)]\end{bmatrix}
\end{aligned} \quad (20)$$

Using the facts shown in [16, 33] that the fourth-order moment of a Gaussian variable is three times the variance squared and that $S(n)$ is symmetric, we get

$$E[A(n)\Delta(n)\Delta(n)A^T(n)] = \begin{bmatrix}(1-2\mu E[f(e(n))]\sigma_x^2 + 2\mu^2 E[f^2(e(n))]\sigma_x^4)S(n) \\ +\mu^2 E[f^2(e(n))]\sigma_x^4 tr[S(n)]I_M\end{bmatrix} \quad (21)$$

$$E[B(n)\delta(n)\delta^T(n)B^T(n)] = \mu^2 E[f^2(e(n))]\sigma_x^4\{\varepsilon(n)\varepsilon^T(n) + tr[\varepsilon(n)\varepsilon^T(n)]I_M\} \quad (22)$$

where $tr(\cdot)$ denotes the trace of a matrix. Also, combining (11), (14), and the definition of $C(\text{n})$, we obtain

$$\begin{aligned}
&E[A(n)\Delta(n)C^T(n)] \\
&= E^T[C(n)\Delta^T(n)A^T(n)] \\
&= (1-\mu E[f(n)]\sigma_x^2)E[W(n)G^T(n)]
\end{aligned} \quad (23)$$

Using (21), (22) and (23), one can derive

$$S(n+1) = \begin{bmatrix} (1-2\mu E[f(e(n))]\sigma_x^2 + 2\mu^2 E[f^2(e(n))]\sigma_x^4)S(n) \\ +\mu^2 E[f^2(e(n))]\sigma_x^4 tr[S(n)]I_M \\ -2\rho(1-\mu E[f(e(n))]\sigma_u^2)E[W(n)C^T(n)] \\ +\mu^2 E[f^2(e(n))]\sigma_x^4(\delta(n)\delta^T(n) + tr[\delta(n)\delta^T(n)]I_M) \\ +\mu^2 E[f^2(e(n))]\sigma_u^2\sigma_v^2 + \rho^2 E[C(n)C^T(n)] \end{bmatrix} \quad (24)$$

Further, the trace of (24) is as follows:

$$tr[S(n+1)] = \begin{bmatrix} (1-2\mu E[f(e(n))]\sigma_x^2 + (M+2)\mu^2 E[f^2(e(n))]\sigma_x^4)tr[S(n)] \\ +(M+1)\mu^2 E[f^2(e(n))]\sigma_x^4\delta(n)\delta^T(n) + M\mu^2 E[f^2(e(n))]\sigma_v^2\sigma_x^2 \\ +\rho^2 E[C(n)C^T(n)] - 2\rho(1-\mu E[f(e(n))]\sigma_x^2)E[W(n)C^T(n)] \end{bmatrix} \quad (25)$$

In equation (24), $\delta(n) = E[\tilde{W}(n)]$, $E[W(n)]$ and $C(n)$ are all bounded, and hence, $E[W(n)C^T(n)]$ converges. This conclusion is the same as that in [16, 33-35] when the sparsity penalty term $g(n)$ are chosen as the $\ell_1$-norm or logarithmic penalty term. The CIM penalty term is also bounded because the negative exponential term can reach its maximum value. Therefore, the adaptive algorithm is stable if the following holds

$$\left|(1-2\mu E[f(e(n))]\sigma_x^2 + (M+2)\mu^2 E[f^2(e(n))]\sigma_x^4\right| < 1. \quad (26)$$

Hence, the above equation (26) reduces to

$$0 < \mu < \frac{2E[f(e(n))]}{(M+2)E[f^2(e(n))]\sigma_x^2} \quad (27)$$

Furthermore, under Assumption 4, we have

$$0 < \mu < \frac{2}{(M+2)E[f(e(n))]\sigma_x^2} = \mu_B$$

Here, we denote the upper bound as the conservative upper bound $\mu_B$ due to Assumption 4.

It is worth noting that the condition of (27) is equal to the stability condition of the standard ZALMS when $e(n) = 0$, or the $\sigma_1 \to \infty$. We also note that $e(n)$ will be approximately equal to $v(n)$ when the proposed algorithm reaches the steady-state (as $n \to \infty$). For this case, (27) can be rewritten as

$$0 < \mu < \frac{2E[f(v(n))]}{(M+2)E[f^2(v(n))]\sigma_x^2} = \frac{2E[\exp(-v^2(n)/2\sigma_1^2)]}{(M+2)E[\exp^2(-v^2(n)/2\sigma_1^2)]\sigma_x^2} \quad (28)$$

This result shows that if the step size satisfies (28), the convergence of the proposed algorithms is guaranteed.

**Remark 4**: According to the mean square convergence analysis, one can see easily that the mean square convergence performance of the CIMMCC algorithm depends mainly on the adaptive filter term. This result is general and coincides with the other sparse adaptive filters proposed in [16, 33].

**Remark 5**: It is worth noting that the condition of (27) is equivalent to the stability condition of the standard ZALMS when $f(e(n))=1$, which is similar to the results in [16, 33-35]. In addition, if the step size satisfies (27), then convergence of the proposed algorithms is guaranteed. However, the right term of (27) is only a conservative upper bound for the proposed algorithm because the condition of (27) is an approximation case. On the one hand, computer simulations, in Section 5, show that the proposed algorithm is converged even if the selected step size is slightly larger than the conservative upper bound. On the other hand, the steady-state performance of the proposed algorithm becomes a tangled mess, and then the result may invaluable for practical engineering. Therefore, the approximate condition in (27) is reasonable for ensuring the algorithm stability.

## 5  Simulation Results

To validate the performance of the proposed CIMMCC algorithm, we adopt state-of-the-art algorithms (i.e., LMP (least mean *p*-power) [21], MCC, ZALMS, RZALMS) and the proposed ZAMCC, RZAMCC algorithms. All the results are obtained by averaging over 100 independent Monte Carlo runs. The parameter vector of the unknown time-varying channel is set as

$$W_o = \begin{cases} [0,0,0,0,0,0,0,0,1,0,0,0,0,0,0,0,0,0] & n \leq 2000 \\ [1,0,1,0,1,0,1,0,1,0,1,0,1,0,1,0,1,0] & 2000 < n \leq 3000 \\ [1,-1,1,-1,1,-1,1,-1,1,-1,1,-1,1,-1,1,-1,1,-1] & 3000 < n \end{cases} \quad (29)$$

In (29), the channel memory size *M* is 20 during different iterations. Here, we define the sparsity degree $D_s$ as:

$$D_s = \frac{N_{non-zero}}{M}$$

where the $N_{non-zero}$ is the number of the non-zero tap in the $W_o$. Since, the channel model has a sparsity of 1/20 during 1 to 2000 iterations, while the $D_s$ changes to 1/2 when the iteration is from 2000 to 3000, and it is non-sparsity after 3000 iterations. Two impulsive noise models including mixed Gaussian (finite variance) and alpha-stable (infinite variance) are considered to in the following computer simulations.

*5.1. Time-varying channel estimation under Mixed Gaussian noise*

We intend to compare the seven algorithms (i..e, LMP, MCC, ZALMS, RZALMS, CIMMCC, ZAMCC, RZAMCC) in the presence of the mixed Gaussian noise. The impulsive observation noise model [41]is defined as

$$(1-\theta)N(\mu_1, \upsilon_1^2) + \theta N(\mu_2, \upsilon_2^2) \quad (30)$$

where $N(\mu_i, \upsilon_i^2)(i=1,2)$ denotes the Gaussian distributions with mean values $\mu_i$ and variances $\sigma_i^2$, and the $\theta$ is the mixture coefficient. In this study, one can notice that the Gaussian distribution with much larger variance can model stronger impulsive noises. The estimation performance is evaluated by mean square deviation（MSD）standard which is defined as

$$MSD(W(n)) = E\{\|W_o - W(n)\|^2\} \quad (31)$$

*Experiment 1*: The convergence behavior of the proposed is first evaluated. Simulation environments are formulated as follows. Parameters $(\mu_1, \mu_2, \upsilon_1, \upsilon_2, \theta)$ in the mixed Gaussian distribution are set as $(0, 0, 0.0001, 20, 0.05)$. $\sigma_1 = 2$ is set as for the standard MCC and the sparse MCC-type algorithms. Step size is $\mu = 0.02$, the zero-attractor controller factors are $\rho = 0.0006$ for sparse aware MCC and $\rho = 0.0001$ for sparse aware LMS. The free parameter $\delta'$ in log-sum penalty for the RZAMCC and RZALMS is 10, and the kernel width in CIMMCC is $\sigma_2 = 0.01$. The parameters are selected such that the algorithms have almost the same initial convergence rate. The convergence curves, in terms of MSD are shown in Fig.1. This figure shows that both ZALMS and RZALMS achieve almost the same convergence performance, and ZAMCC and RZAMCC also obtain similar performance. Unlike these algorithms, the CIMMCC achieve better performance during the first and second stages. After 3000 iterations, the CIMMCC algorithm performs comparable with the standard MCC and other sparse MCC-type algorithms even in the case of the non-sparse system. Hence, our proposed algorithm can well track time-varying channels.

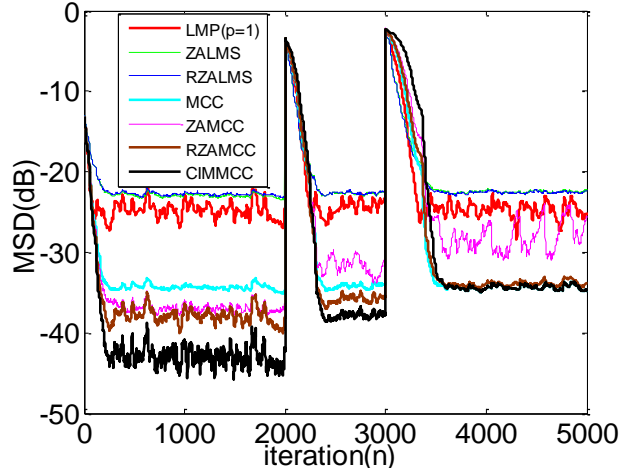

Fig.1. Tracking and steady-state behaviours of 20-order adaptive filters,

driven by white input signal and corrupted by mixed Gaussian noise.

*Experiment 2*: We further examine MCC aware algorithms with different step sizes (i.e., 0.005, 0.01, 0.05, 0.085, 0.1 and 0.5). The simulation results are depicted in Fig.2. One can see that the performance of the proposed algorithm becomes better with smaller and vice versa. In practice, an upper bound of the step size is needed to guarantee the convergence performance of the MCC and sparse aware MCC algorithms. In this example the optimal performance is achieved at the step size 0.005 for all algorithms. Furthermore, the performance curves of the sparse MCC algorithms are almost the same as the standard MCC algorithm when step size gets larger, though the performance becomes poorer. Simulation results suggest that the step sizes of the sparse MCC algorithms need unifying upper bound to ensure the convergence performance, as proven in section 4. In this case, the conservative upper bound of $\mu$ obtained from (27) is 0.085. We observe that the satisfactory convergence performance can be achieved when $\mu = 0.05 < \mu_B$. Since $\mu_B$ is a conservative upper bound, the algorithm still show the convergence behavior when the step size is slightly larger than $\mu_B$ (e.g. $\mu = 0.1$), but the performance become much poorer which doesn't meet the requirement of practical engineering.    When the step size is too large (say $\mu > 0.1 > \mu_B$), the algorithm shows no obvious convergence. The error bars of the CIMMCC in terms of MSD are plotted in Fig.3 to further validate the results. According to the result in Fig.3, one can observe that the MSD of the CIMMCC is very worse when the step size $\mu$ is larger than 0.085.

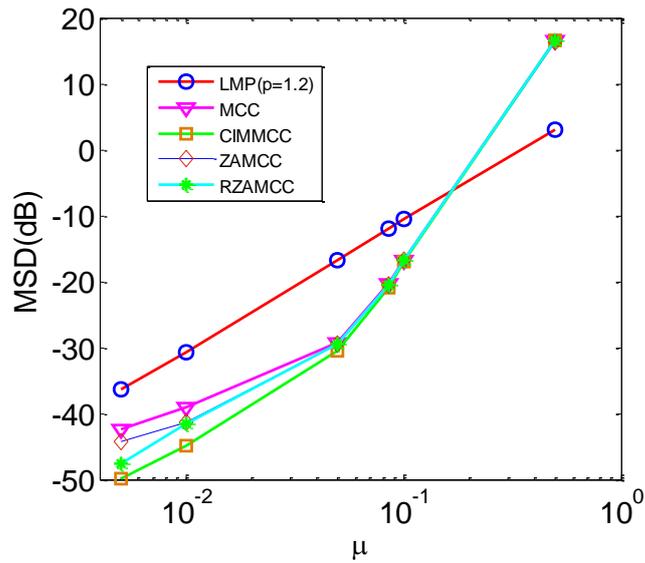

Fig.2. Steady-state MSDs of the channel estimates versus step sizes

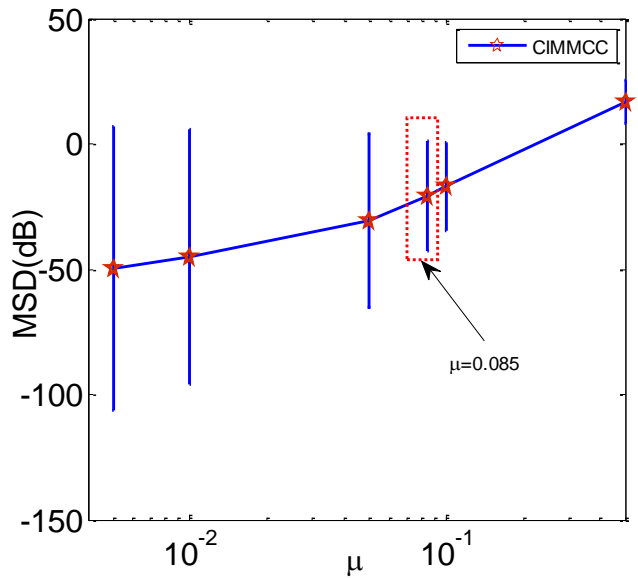

Fig.3. Error bar performance of CIMMCC with different step sizes

*5.2 Time-varying channel estimation under Alpha -stable noise*

The mixed Gaussian models do not suitable for modeling all of natural noises [38].Even though its tail decays exponentially while empirical evidence manifests that algebraic decay of heavy-tailed noise processes often occurs in communications as well as other fields [39]. To solve this problem, we consider the alpha-stable noise model which provides a good description for above mentioned heavy-tailed noises [41].Hence, alpha-stable model has been attracted highly attentions [20-22]. Under the alpha-stable noises, we also verify the effectiveness of the proposed CIMMCC algorithm.

First of all, the characteristic function of alpha-stable process is given by

$$f(t) = \exp\{j\delta t - \gamma |t|^\alpha [1 + j\beta \operatorname{sgn}(t) S(t,\alpha)]\} \tag{32}$$

in which

$$S(t,\alpha) = \begin{cases} \tan\dfrac{\alpha\pi}{2} & if\ \alpha \neq 1 \\ \dfrac{2}{\pi}\log|t| & if\ \alpha = 1 \end{cases} \tag{33}$$

where $\alpha \in (0,2]$ is the characteristic factor, $-\infty < \delta < +\infty$ is the location parameter, $\beta \in [-1,1]$ is the symmetry parameter, and $\gamma > 0$ is the dispersion parameter. The characteristic factor $\alpha$ measures the tail heaviness of the distribution. Smaller $\alpha$ is means to the heavier tail in (31). In addition, $\gamma$ measures the dispersion of the distribution, which plays a similar role to the variance of Gaussian distribution. The distribution is symmetric about its location $\delta$ when $\beta = 0$. Such a distribution is called a symmetric alpha-stable distribution ($S\alpha S$). In our simulations, the noise model $v(n)$ is defined as $V = (\alpha, \beta, \gamma, \delta)$.

***Experiment 1***: Let us consider the problem of sparse channel estimation under impulsive noise model with $V = (1.2, 0, 0.2, 0)$. The step size is set at 0.02 for all of adaptive filtering algorithms: LMP, MCC, ZALMS, RZALMS, ZAMCC, RZAMCC, and CIMMCC. The kernel widths in MCC and CIM are set as 2.0 and 0.01, respectively. For ZALMS, RZALMS, ZAMCC, RZAMCC, and CIMMCC, the weight factor of the sparsity inducing term is set at 0.0001 and $\delta' = 10$. The average convergence curves in terms of the mean square deviation (MSD) are shown in Fig.4. One can find that sparse MCC-type algorithms achieve faster convergence rate and better steady-state performance than the standard algorithms (i.e., LMP and MCC). In addition, Fig. 4 also shows that proposed CIMMCC achieves lower MSD than

ZAMCC and RZAMCC because the CIM provides a better approximation to $\ell_0$-norm function. It is worth notice that both ZALMS and RZALMS are unstable which is caused by impulsive noises. In the subsequent experiments, unstable LMS-type algorithms are omitted and only stable algorithms (LMP and MCC) as the benchmark will be compared.

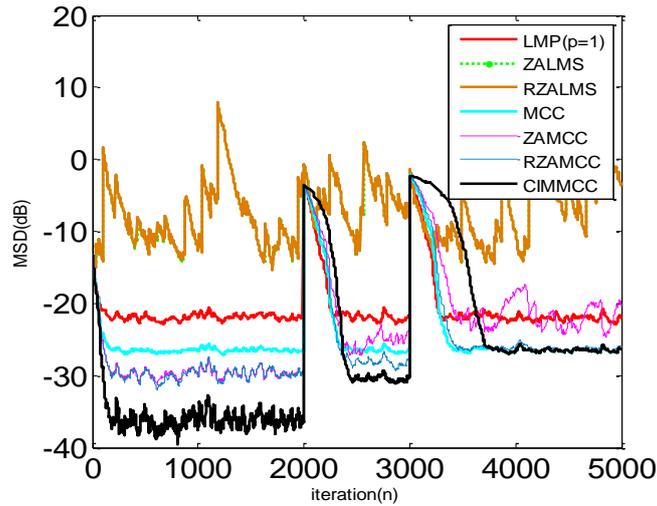

Fig.4.Tracking and steady-state behaviours of 20-order adaptive filters, driven by white input signal corrupted by impulsive noise

*Experiment 2*:To further demonstrate the performance of the proposed methods, we continue to conduct the simulation with different values of the different $\gamma$ (1,1.5,2,2.5,3) and different $\alpha$ (0.5,1,1.1,1.2,1.3,1.4,1.5,1.6,1.7) inFig.5and Fig. 6. As shown in the two figures, the curve of our proposed CIMMCC is lower than other algorithms.

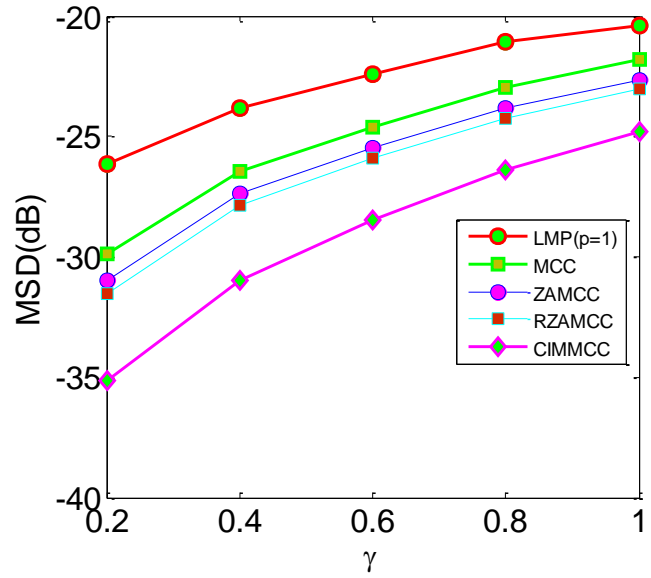

Fig.5.MSD of the channel estimates versus $\gamma$

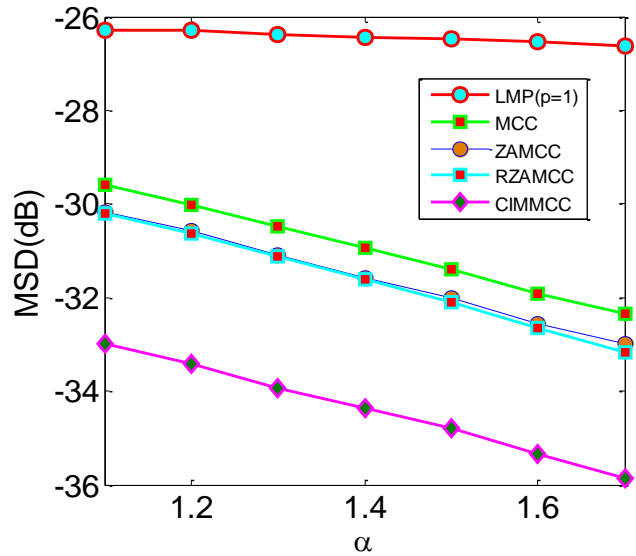

Fig.6.MSD of the channel estimates versus $\alpha$.

**Experiment 3**: The kernel size $\sigma_1$ in MCC is one of important parameters for the CIMMCC. To study the connection between the parameter and CIMMCC, MSD curves of CIMMCC are depicted in different $\sigma_1$ as shown in Fig.7. This figure implies that the proposed algorithm under less impulsive noise (bigger α) can achieve the better performance for different $\sigma_1$. Hence, the simulation result in

Fig. 7is also coincidence with theory analysis in Section 4. In addition, the lower MSD of the proposed algorithm is obtained when $\sigma_1 = 1$ for different alpha. Hence, the MSD is deteriorating as the kernel size increasing in exceeds certain range. For obtaining the optimal performance, it is worth noting that suitable choosing the kernel size is also important for the CIMMCC.

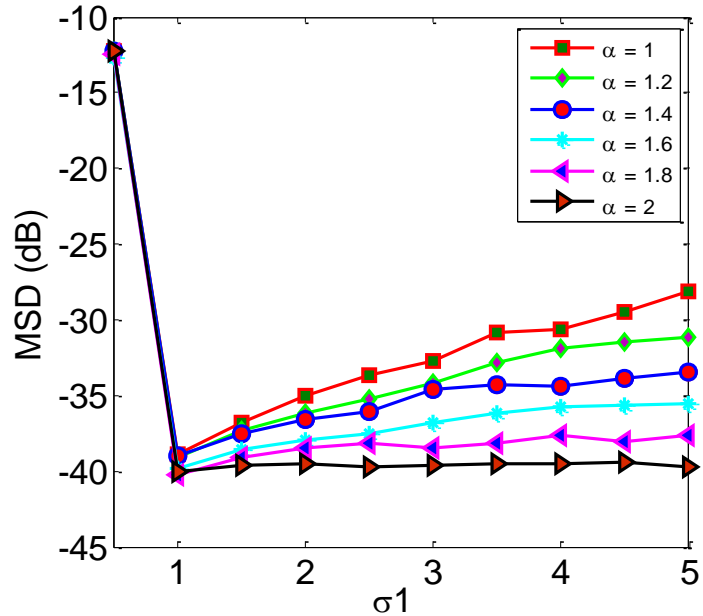

Fig.7.MSD of CIMMCC with different kernel size for different $\alpha$.

*Experiment 4*: The effects of the step size on the performance of the CIMMCC algorithm will be studied as follows. Proper step size ensures good convergence performance. We illustrate the convergence behaviors of the CIMMCC with different step sizes. The noise parameters is set as $V = (1.4, 0, 1, 0)$. The kernel width $\sigma_2$ is 0.05, the zero-attract factor is $\rho = 0.0005$, and the *p* value is *p*=1.2. The convergence curves viruses step sizes (0.01, 0.03, 0.05, 0.07, 0.09, 0.1, 0.2, and 0.3) are depicted in Fig.8. One can observe that the satisfactory convergence performance is achieved when $\mu = 0.01$. However, when the step size is too large (i.e., $\mu > 0.2$), the algorithm does not converge. The error bar performance is illustrated as well in Fig.9. To keep the algorithm converges to a steady-state value, $\mu$ should be set a conservative upper bound. Hence, the simulation result is coincidence with the convergence analysis in section 4.

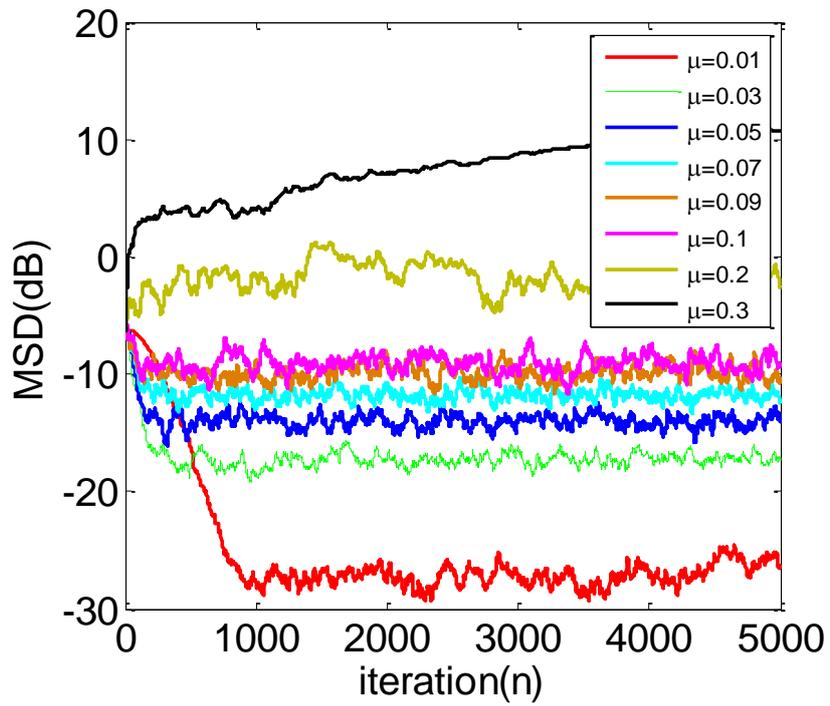

Fig.8. Convergence of the CIMMCC with different $\mu$

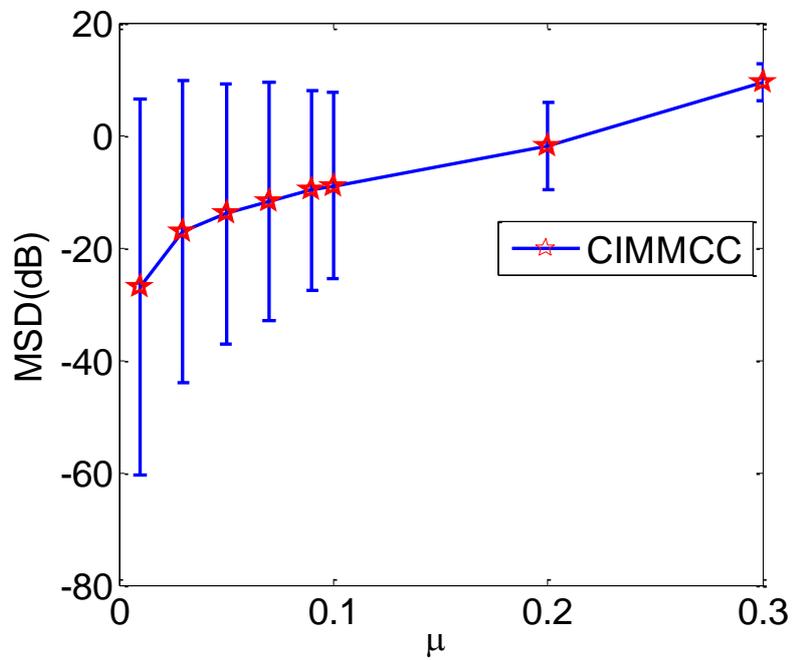

Fig.9. Error bar performance of the CIMMCC with different step sizes

*5.3 Sparse echo cancellation under alpha stable noise*

One of the important applications of sparse adaptive system identification is the sparse echoes cancellation [42-44].In this section, we evaluate the ability of the proposed sparse aware MCC algorithms using sparse echo cancellation scenarios. In our simulation, the echo path commutes to a channel with length $M$ = 1024 and 52 non-zero coefficients. This artificial echo path, which has been represented in Fig.10, will be referred as the sparse echo channel.

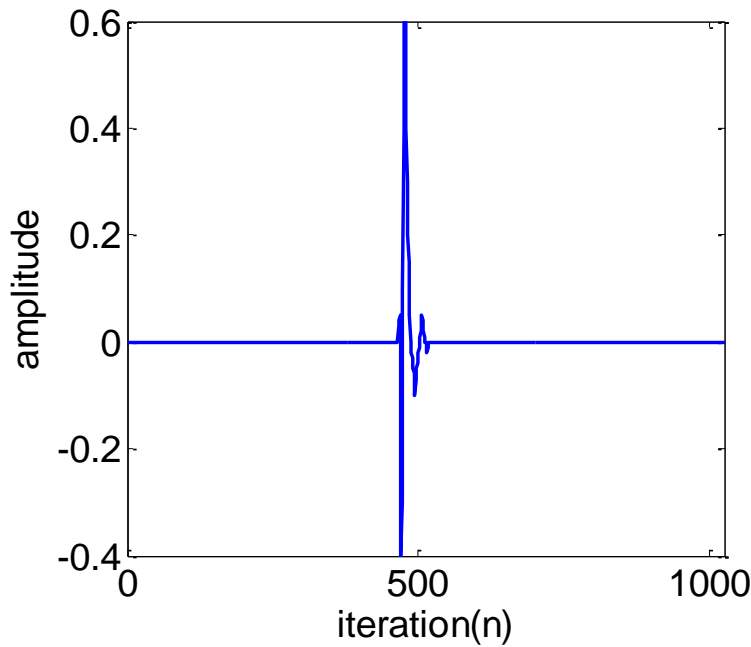

Fig.10. Impulse response of artificially generated sparse echo channels

***Experiment 1***: USASI noise [45] input with a speech-like spectrum [46] and the alpha-stable noise model are considered in the simulation. $\sigma_1$ and $\sigma_2$ are set at 2 and 0.01, respectively; the zero-attract factor is $\rho = 0.0001$, and the step size are set as 0.001 and 0.0005, $\delta' = 10$ is considered. Fig.11 shows the convergence curves for MCC-type algorithms, i.e., CIMMCC, RZAMCC as well ZAMCC. It is expected that all the sparse MCC-type algorithms converge quickly and the proposed CIMMCC algorithm achieve the optimal MSD performance.

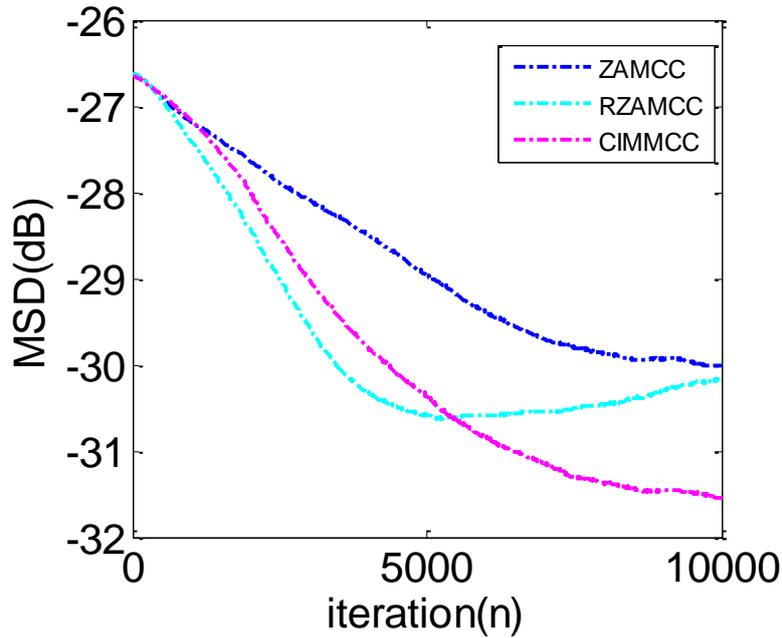

Fig.11. Convergence for sparse echo response with USASI noise input

***Experiment 2***: To conclude the evaluation of the proposed schemes for echo cancellation further, we demonstrated their performance when the input signal is a fragment of 2s of real speech, sampled at 8 kHz in this subsection. Simulation parameters are set as: $\rho = 0.0001$, $\mu = 0.001$, $\sigma_1 = 2$, $\sigma_2 = 0.01$, $\delta' = 10$ and the simulation result is shown in Fig.12. This figure shows that the proposed sparse MCC algorithms perform well performance for the real speech under the impulsive noise case. According to the simulation, one can deduce that the proposed MCC algorithms can work well in the practical scenarios of sparse echo cancellation.

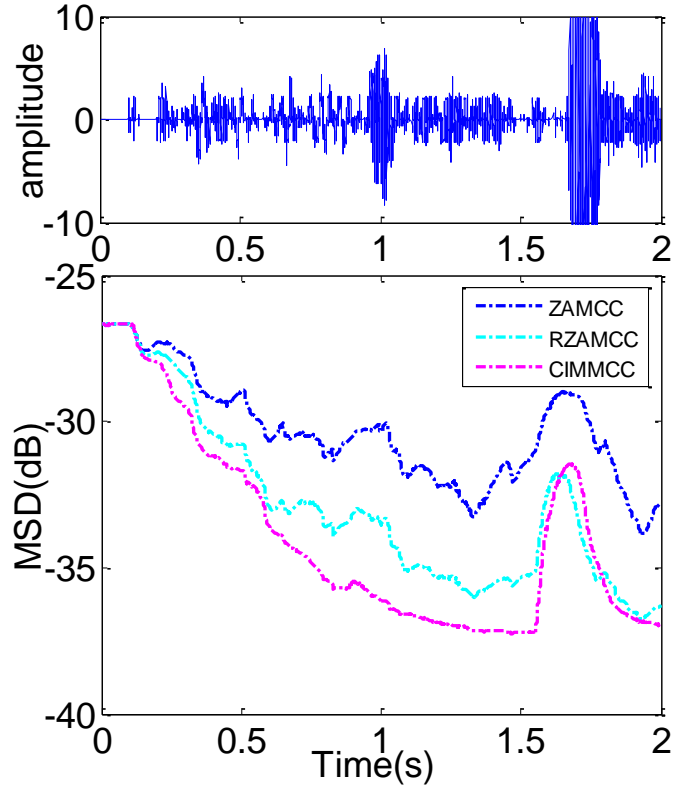

Fig.12. Convergence for sparse echo response with speech input.

## 6 Conclusions

In this paper, we proposed anew robust and sparse adaptive filtering algorithm, derived by incorporating into the MCC with a CIM based sparsity penalty term. We have analyzed theoretically the mean and mean square convergence of the proposed algorithm, and derived certain conditions for guaranteeing convergence. Simulation results confirmed the desirable performance of the new algorithm under impulsive noise environments.

## Acknowledgements

This work was supported by National Natural Science Foundation of China (NSFC) grants (No. 61371807, No. 61372152, No.61401069), 973 Program (2015CB351703), and Key Project of major national science and technology on new generation of broadband wireless mobile communication network (2012ZX03001023-003, 2012ZX03001008-003, and 2013ZX03002010-003) as well as

Japan Society for the Promotion of Science (JSPS) research activity start-up research grant (No. 26889050).

## *Appendix 1*:

1. Sparse MCC with *zero-attracting ($\ell_1$-norm) penalty term(ZAMCC)*

To develop a sparse MCC algorithm with zero-attracting ($\ell_1$-norm) penalty term, we introduce the cost function

$$J_{CIMMCC}(W(n)) = -J_{MCC}(n) + \lambda J_{ZA}(W(n))$$
$$= -\frac{1}{\sigma_1\sqrt{2\pi}}\exp\left(-e^2(n)/2\sigma_1^2\right) + \lambda \|W(n))\|_1 \quad (34)$$

where $J_{ZA}(W(n)) = \|W(n))\|_1$ denotes the $\ell_1$-norm of the estimated parameter vector. In (34), the MCC term is robust to impulsive noise, and the ZA penalty term is a sparsity inducing term, and the two terms are balanced by a weight factor $\lambda \geq 0$. Based on the cost function (34), we derive an adaptive algorithm:

$$W(n+1) = W(n) - \eta \frac{\partial J_{ZAMCC}(W(n))}{\partial W(n)}$$
$$= W(n) - \eta \left[\frac{-1}{\sigma_1^3\sqrt{2\pi}}\exp(-e^2(n)/2\sigma_1^2)e(n)X(n) + \lambda sign(W(n))\right] \quad (35)$$
$$= W(n) + \mu\exp(-e^2(n)/2\sigma_1^2)e(n)X(n) - \rho sign(W(n))$$

where $\mu$ and $\rho$ are the same as in (8) and (9), and the $sign(\cdot)$ is a component-wise sign function. This algorithm is referred to as the ZAMCC algorithm.

2. Sparse MCC with the logarithmic penalty term

In this part, we derive a sparse MCC algorithm with logarithmic penalty term. We define the following cost function:

$$J_{RZAMCC}(W(n)) = -J_{MCC}(W(n)) + \lambda J_{RZA}(W(n))$$
$$= -\frac{1}{\sigma_1\sqrt{2\pi}}\exp\left(-e^2(n)/2\sigma_1^2\right) + \lambda\sum_{i=1}^{M}\log\left(1+|w_i|/\delta\right) \quad (36)$$

where the log-sum penalty $\sum_{i=1}^{M}\log(1+|w_i|/\delta)$ is introduced as it behaves more similarly to the

$\ell_0$-norm than $||W||_1$, $\delta$ is a positive number. Then a gradient based adaptive algorithm can be easily derived as

$$w_i(n+1) = w_i(n) - \eta \frac{\partial J_{RZAMCC}(W(n))}{\partial w_i(n)}$$
$$= w_i(n) - \eta \left[ \begin{array}{c} \frac{-1}{\sigma_1^3 \sqrt{2\pi}} \exp(-e^2(n)/2\sigma_1^2) e(n) x_i(n) \\ + \lambda \frac{\text{sign}(w_i(n))}{1+\delta'|w_i(n)|} \end{array} \right] \quad (37)$$
$$= w_i(n) + \mu \exp(-e^2(n)/2\sigma_1^2) e(n) x_i(n) - \rho \frac{\text{sign}(w_i(n))}{1+\delta'|w_i(n)|}$$

or equivalently, in vector form

$$W(n+1) = W(n) + \mu \exp\left(-\frac{e^2(n)}{2\sigma_1^2}\right) e(n) X(n) - \rho \frac{\text{sign}(W(n))}{1+\delta'|W(n)|} \quad (38)$$

where $\mu$ and $\rho$ are the same as in (8) and (9), and $\delta' = 1/\delta$. This algorithm is referred to as the RZAMCC algorithm.

## *Appendix 2*:

We define a function as follows:

$$\bar{g}(x) = \tau x \exp\left(-x^2/2\sigma^2\right) \quad (39)$$

where $\tau$ is a constant. Obviously, we have

$$\bar{g}(x) = g_{CIM}(w_i(n)) = -\frac{1}{M\sigma^3\sqrt{2\pi}} w_i(n) \exp(-w_i(n)^2/2\sigma^2) \quad (40)$$

when $\tau = -1/(M\sigma^3\sqrt{2\pi})$ and $x = w_i(n)$. Further, we compute the limit of the function $\bar{g}(x)$ based on L'Hôpital's rule with $x \to \infty$

$$\lim_{x \to \infty} \bar{g}(x) = \lim_{x \to \infty} \tau x \exp\left(-\frac{x^2}{2\sigma^2}\right) = \tau \lim_{x \to \infty} \frac{x}{e^{\frac{x^2}{2\sigma^2}}} = \tau \lim_{x \to \infty} \frac{\sigma^2}{x e^{\frac{x^2}{2\sigma^2}}} = 0 \quad (41)$$

The estimated weight $w_i(n)$ is limited in general, and by (41) even if the channel parameter $w_i(n)$ tends to infinity, the vector $g_{CIM}(w_i(n))$ is still limited.